\documentclass[notitlepage,a4,10.5pt]{article}

\usepackage{graphicx}

\usepackage{amsmath}%
\usepackage{amsfonts}%
\usepackage{amssymb}%
\usepackage{mathrsfs}
\usepackage{amsthm}
\usepackage{float}

\usepackage{authblk}

\usepackage[left=2.5cm,right=2.5cm,top=3cm,bottom=3cm]{geometry} 

\usepackage{xcolor}
\usepackage{comment}

\newcommand{\const}{{\mathrm{const}}}

\newcommand{\mc}{\mathcal}
\newcommand{\cp}{\times}
\newcommand{\bol}{\boldsymbol}

\newcommand{\abs}[1]{\left\lvert{#1}\right\rvert}

\newcommand{\lr}[1]{\left({#1}\right)}
\newcommand{\lrs}[1]{\left[{#1}\right]}
\newcommand{\lrc}[1]{\left\{{#1}\right\}}

\newcommand{\p}{\partial}
\newcommand{\ti}[1]{\textit{#1}}
\newcommand{\tb}[1]{\textbf{#1}}

\newcommand{\fdv}[2]{\frac{\delta{#1}}{\delta{#2}}}
\newcommand{\fdvdv}[3]{\frac{\delta^2{#1}}{\delta{#2}\,\delta{#3}}}
\newcommand{\sumcyc}{\sum_{F,G,H: \text{cyclic}}}

\newcommand{\dd}{\mathrm{d}}
\newcommand{\vrho}{\varrho_0}
\newcommand{\gmr}{\gamma\vrho}

\newcommand{\tlaplacian}{{\Delta}_{\gamma}}

\newcommand{\R}{\mathbb{R}}

\newcommand{\tagref}[1]{\tag{\ref{#1}}}

\newcommand{\eq}[1]{\begin{equation}\begin{split}{#1}\end{split}\end{equation}}

\begin{document}

\title{
Revealing Noncanonical Hamiltonian Structures \\in Relativistic Fluid Dynamics}
\author[1,2]{Keiichiro Takeda} \author[2]{Naoki Sato}
\affil[1]{Graduate School of Frontier Sciences, The University of Tokyo, \protect\\ 5-1-5 Kashiwa-no-ha, Kashiwa-city, Chiba 277-8561, Japan\protect\\ Email: takeda.keiichiro20@ae.k.u-tokyo.ac.jp}
\affil[2]{National Institute for Fusion Science, \protect\\ 322-6 Oroshi-cho Toki-city, Gifu 509-5292, Japan \protect\\ Email: 
sato.naoki@nifs.ac.jp}
\date{\today}
\setcounter{Maxaffil}{0}
\renewcommand\Affilfont{\itshape\small}

    \maketitle
    \begin{abstract}
We present the noncanonical Hamiltonian structure of the relativistic Euler equations for a perfect fluid in Minkowski spacetime. By identifying the system’s noncanonical Poisson bracket and Hamiltonian, we show that relativistic fluid flows preserve helicity and enstrophy as conserved quantities in three-dimensional and two-dimensional cases, respectively. This holds when the fluid follows a relativistic $\gamma$-barotropic equation of state, which generalizes the classical barotropic condition. Furthermore, we demonstrate that these conserved quantities are Casimir invariants associated with the noncanonical Poisson structure. These findings open new avenues for applying Hamiltonian theory to the study of astrophysical fluids and relativistic plasmas. 
    \end{abstract}



\section{Introduction}
Invariants play a crucial role in understanding fluid dynamics \cite{Arnold2009}, particularly in the study of turbulent flows. In three-dimensional ideal flows, helicity \cite{Moffatt14,Enciso2016helicity}—a scalar quantity that measures the linkage and twisting of vortex lines—is conserved, imposing significant constraints on the topology of vortex configurations. In contrast, in two-dimensional ideal flows, enstrophy, defined as the \(L^2\) norm of vorticity, serves as an invariant. Each of these invariants influences fluid motion in distinct ways, shaping the evolution of vortex structures \cite{Enciso2014topological} and affecting the distribution and transfer of energy across different scales. Notably, in two-dimensional turbulence, the presence of enstrophy conservation leads to the inverse energy cascade \cite{Kraichnan1967inertial,Kraichnan1980two-dimensional}, a phenomenon where energy flows from smaller to larger scales, in stark contrast to the behavior observed in three-dimensional turbulence \cite{K41}.

Interestingly, both helicity and enstrophy emerge as \ti{Casimir invariants} within the noncanonical Hamiltonian framework for ideal fluid motion. This arises when the {baroclinic term} \( \nabla \rho \times \nabla P \), representing the departure from a barotropic equation of state \( P(\rho) \), vanishes in the vorticity equation. Here, \( \rho \) is the fluid mass density and \( P \) the pressure \cite{Morrison98,Morrison80}.
 A Casimir invariant is a quantity that lies in the kernel (null space) of the Poisson bracket, which defines the noncanonical Hamiltonian structure of the system \cite{Morrison82,Littlejohn82,Olver1982}. Crucially, this means that Casimir invariants are conserved quantities, remaining unchanged under the evolution of the system, irrespective of the specific form of the Hamiltonian (or energy function). This property underscores their fundamental role in governing the dynamics of the fluid.

A key issue, particularly in the study of astrophysical fluids, is whether the properties of ideal flows discussed above extend to the framework of special relativity. It is well-established that ideal magnetohydrodynamics (MHD)—which generalizes the Euler equations to include the effects of magnetic fields—admits Hamiltonian formulations \cite{Abdelhamid2015} that are compatible with special relativity (see e.g., \cite{DAvignon2015,Holm1986,Kawazura2017}). Furthermore, both helicity \cite{Yoshida2014} and enstrophy \cite{Nunotani2022} have been successfully extended to special relativistic ideal fluid flows. In these relativistic formulations, helicity and enstrophy are defined in terms of the four-momentum, utilizing intrinsically four-dimensional quantities, and are conserved in a co-moving domain with respect to proper time.

However, two important questions remain unresolved: (i) whether it is possible to define helicity and enstrophy in terms of the three-dimensional velocity field associated with spatial coordinates and ensure their conservation with respect to the classical time variable under a special equation of state generalizing the barotropic condition $P\lr{\rho}$ to the relativistic regime, and (ii) whether these quantities correspond to Casimir invariants within a noncanonical Hamiltonian framework. In this work, we demonstrate that both of these questions have affirmative answers for special relativistic flows governed by a \ti{$\gamma$-barotropic} equation of state.

These findings pave the way for novel applications of Hamiltonian theory to the study of astrophysical fluids and relativistic plasmas. For instance, by employing an appropriate Clebsch representation of the velocity field \cite{YosClebsch,Yoshida2017}, the noncanonical Hamiltonian structures identified in this work can be extended to formulate diffusion \cite{Sato21} and collision operators \cite{Morrison24,Sato24}. These operators would account for dissipative effects and relaxation dynamics while adhering to the constraints imposed by Casimir invariants, such as relativistic helicity and enstrophy. This framework provides a powerful tool for modeling the evolution of fluid systems in space environments, where conservation laws play a critical role in shaping dynamics.

The structure of this paper is organized as follows: {Section 2} reviews the derivation of the relativistic Euler equations, starting from the energy-momentum tensor of a perfect fluid. {Section 3} demonstrates that these flows possess a noncanonical Hamiltonian structure. In {Section 4}, we introduce the concept of a \(\gamma\)-barotropic equation of state, which represents one of the possible generalizations of the classical barotropic condition \(P(\rho)\) to the relativistic framework.

{Sections 5 and 6} explore how the Hamiltonian structure, developed in Section 3, simplifies under the \(\gamma\)-barotropic equation of state. We show that the resulting equations preserve relativistic helicity in three dimensions and relativistic enstrophy in two dimensions. These invariants are expressed in terms of the spatial velocity and vorticity fields and remain conserved with respect to the classical time variable. Additionally, they naturally reduce to the classical invariants when the Lorentz factor \(\gamma\) approaches 1.

In {Section 7}, we discuss how the conservation of helicity and enstrophy is violated if the \(\gamma\)-barotropic condition is relaxed, due to the emergence of a \(\gamma\)-baroclinic effect in the vorticity equation. {Section 8} concludes the paper with final remarks and a discussion of future directions.

\section{Relativistic Euler Equations}
In this section, we begin by revisiting the derivation of the relativistic Euler equation, originating from the energy-momentum tensor of a perfect fluid within a relativistic framework. 

Let $u^\mu\lr{\bol{x},t}$, $\mu=0,...,3$, $\rho_0\lr{\bol{x},t}$, and $P\lr{\bol{x},t}$ represent the components of the four-velocity, the mass density as measured by a co-moving observer with velocity $u$, and the pressure field, respectively. Here, $\lr{\bol{x},t}\in\Omega\times [0,\infty)$, where $\Omega\subset\mathbb{R}^3$ is a three-dimensional smooth bounded domain with boundary $\p\Omega$. 

The energy-momentum tensor of a perfect fluid~\cite{Frankel79} is given by 
\begin{equation}
T^{\mu\nu}=\lr{\rho_0+\frac{P}{c^2}}u^{\mu}u^{\nu}+P\eta^{\mu\nu}.
\end{equation}
Here, $\eta^{\mu\nu}$ is the Minkowski metric tensor with signature $\lr{-,+,+,+}$.  
The equations of motion are derived from the divergence of the energy-momentum tensor, 
\begin{equation}
\p_{\mu}T^{\mu\nu}=
\p_{\mu}\lrs{\lr{\rho_0+\frac{P}{c^2}}u^{\mu}u^
{\nu}+P\eta^{\mu\nu}}=0. \label{EoM}
\end{equation}
The four-velocity $u$ can be expressed as 
\begin{equation}
u=u^0\lr{\bol{x},t}\p_0+u^1\lr{\bol{x},t}\p_1+u^2\lr{\bol{x},t}\p_2+u^3\lr{\bol{x},t}\p_3 .
\end{equation}
Normalization 
\eq{
u^{\mu}u_{\mu}=-c^2,
}
leads to the relation
\begin{equation}
u_0^2=c^2+\bol{u}^2,
\end{equation}
with $\bol{u}=u^1\p_1+u^2\p_2+u^3\p_3$ and  $\bol{u}^2=\abs{\bol{u}}^2=u_1^2+u_2^2+u_3^2$.
In the following, we take the root  $u^0=-u_0=\sqrt{c^2+\bol{u}^2}$. 
Thus, system \eqref{EoM} can be reduced to the following system of equations for the unknowns $u^1$, $u^2$, $u^3$, 
\eq{\vrho=\rho_0+\frac{P}{c^2},} and $P$,  
\begin{subequations}
\begin{align}
&\p_{\mu}\lr{\vrho\,u^{\mu}u^0}=\p_0 P \label{EoM1-0}\\
&\p_{\mu}\lr{\vrho\,u^{\mu}u^1}=-\p_1 P \\
&\p_{\mu}\lr{\vrho\,u^{\mu}u^2}=-\p_2 P \\
&\p_{\mu}\lr{\vrho\,u^{\mu}u^3}=-\p_3 P
.
\end{align}
\end{subequations}
Setting $\lr{x^0,x^1,x^2,x^3}=\lr{ct,x,y,z}$ and denoting with $\nabla$ the usual gradient operator in $\R^3$, it can be shown (see Appendix \ref{Derivation of Equation (EoM2-0)}) that the system above is equivalent to 
\begin{subequations}
\begin{align}
&\frac{c^3}{u^0}\lrs{\frac{\p}{\p t}\lr{\vrho\,\sqrt{1+\frac{\bol{u}^2}{c^2}}}+\nabla\cdot\lr{\vrho\,\bol{u}}}=\frac{\p P}{\p t}+\frac{\bol{u}\cdot\nabla P}{\sqrt{1+\frac{\bol{u}^2}{c^2}}} , \label{EoM2-0}\\
&\lrs{\frac{\p}{\p t}\lr{\vrho\,\sqrt{1+\frac{\bol{u}^2}{c^2}}}+\nabla\cdot\lr{\vrho\,\bol{u}}}\bol{u}+\vrho\lr{\sqrt{1+\frac{\bol{u}^2}{c^2}}\frac{\p\bol{u}}{\p t}+\bol{u}\cdot\nabla\bol{u}}=-\nabla P.
\end{align}\label{EoM2}
\end{subequations}
Noting that $\vrho=\rho_0+P/c^2$ plays the role of the mass density,
where $P/c^2$ physically encodes the mass density associated with thermal fluctuations, we impose the following continuity equation for the variable $\vrho$, 
\begin{equation}
\frac{\p}{\p t}\lr{\vrho\,\sqrt{1+\frac{\bol{u}^2}{c^2}}}+\nabla\cdot\lr{\vrho\,\bol{u}}=0,
\end{equation}
which must be solved together with 
\begin{subequations}
\begin{align}
&\frac{\p P}{\p t}+\frac{\bol{u}\cdot\nabla P}{\sqrt{1+\frac{\bol{u}^2}{c^2}}}=0,\\
&\sqrt{1+\frac{\bol{u}^2}{c^2}}\frac{\p\bol{u}}{\p t}+\bol{u}\cdot\nabla\bol{u}=-\frac{1}{\vrho}\,\nabla P.
\end{align}\label{EoM3}
\end{subequations}
Recall that the proper time $s$ is defined as 
\eq{-c^2ds^2=-c^2dt^2+dx^2+dy^2+dz^2.} 
Since the Lorentz factor is given by the ratio  $\gamma=dt/ds$, we find 
\eq{\gamma^2=1+\frac{\bol{u}^2}{c^2},} so that the five governing equations for the five unknowns $\vrho$, $P$, and $\bol{u}$ can be written as
\begin{subequations} \label{EoM4}
\begin{align}
&\frac{\p}{\p t}\lr{\gmr}+\nabla\cdot\lr{\vrho\,\bol{u}}=0, \label{EoM4-theta}\\
&\gamma\frac{\p P}{\p t}+{\bol{u}\cdot\nabla P}=0, \label{EoM4-P}\\
&\gamma\frac{\p\bol{u}}{\p t}+\bol{u}\cdot\nabla\bol{u}=-\frac{1}{\vrho}\,\nabla P \label{EoM4-v}.
\end{align}
\end{subequations}
System \eqref{EoM4} encapsulates the relativistic Euler equations governing the dynamics of a perfect fluid.
In the non-relativistic limit, where \( c \) is sufficiently large, the right-hand side of equation (\ref{EoM2-0}) becomes negligible. As a result, equation (\ref{EoM4-P}) no longer plays a role in the non-relativistic Euler equations, leaving only the classical continuity and momentum equations to describe the system.

\section{Noncanonical Hamiltonian Structure of a Perfect Fluid}

In this section, we derive the Hamiltonian structure, including both the Hamiltonian functional and the Poisson operator, for a perfect fluid without assuming a specific equation of state that relates pressure to the other dynamical variables.

The phase space of the system is spanned by the triplet $\bol{z}=\lrc{\bol{u},\gmr,P}$.
Note that we take as independent variables the triplet $\lrc{\bol{u},\gmr,P}$, and not  $\lrc{\bol{u},\vrho,P}$.

The Hamiltonian for a perfect fluid, representing the total energy of the system, is given by:
\begin{equation}
\mc{H}\lrs{\bol{u},\gmr,P}=\int_\Omega\lrs{c^2\gamma^2\vrho-P}\dd^3x
=\int_\Omega\lrs{c\gmr\sqrt{c^2+\bol{u}^2}-P}\dd^3x .\label{H1}
\end{equation}
This quantity is conserved, as can be shown by applying the equations of motion \eqref{EoM4} and the boundary condition:
\eq{
\gmr\bol{u}\cdot\bol{n}=0~~~~{\rm on}~~\p\Omega. \label{un}
}
Indeed, we have:  
\begin{equation}
\frac{\dd\mc{H}}{\dd t}
=-\int_\Omega\nabla\cdot\lr{c^2\gmr\bol{u}}\dd^3x.
\end{equation}

In what follows, we consider functionals \( F[\boldsymbol{u}, \gamma \rho, P] \) that satisfy the boundary condition:
\begin{equation}
    \frac{\delta F}{\delta \boldsymbol{u}}\cdot\boldsymbol{n} = 0 \quad \text{on} \quad \partial \Omega,
    \label{gen3D-functionals-boundary}
\end{equation}
where \( \boldsymbol{n} \) represents the unit outward  normal to the bounding  surface \( \partial \Omega \). These boundary conditions are necessary to ensure the validity of the Poisson bracket axioms, particularly the antisymmetry property and the Jacobi identity.

Next, notice that the functional derivatives of $\mc{H}$ are 
\begin{equation}
\fdv{\mc{H}}{\bol{u}}=\vrho\bol{u},~~~~
\fdv{\mc{H}}{\lr{\gmr}}=c^2\gamma,~~~~
\fdv{\mc{H}}{P}=-1 .
\end{equation}
Hence, the equations of motion \eqref{EoM4} can be expressed in the form  
\eq{
\frac{\dd\bol{z}}{\dd t}=\lrc{\bol{z},\mc{H}},
}
where the Poisson bracket $\lrc{\cdot,\cdot}$ assigns the evolution 
of an observable $F$ according to  
\eq{
\frac{\dd F}{\dd t}=\lrc{F,\mc{H}}=\int_{\Omega}\frac{\delta F}{\delta\bol{z}}\cdot\mc{J}\,\frac{\delta\mc{H}}{\delta\bol{z}}\,\dd^3x.
}
Here, $\mc{J}$ denotes the Poisson operator
\begin{equation}
\mc{J}=\begin{bmatrix}
-\frac{1}{\gmr}\,\bol{\omega}\cp & -\nabla & \frac{1}{\gmr}\,\nabla P \\
-\nabla\cdot & 0 & 0 \\
-\frac{1}{\gmr}\nabla P\cdot & 0 & 0
\end{bmatrix}.
\end{equation}
Explicitly, the Poisson bracket can be written as 
\begin{align}
\lrc{F,G}
&=\int_\Omega\dd^3 x\,\lrs{\fdv{F}{\bol{u}}\cdot\lr{-\frac{\bol{\omega}}{\gmr}\cp\fdv{G}{\bol{u}}-\nabla\fdv{G}{\lr{\gmr}}+\frac{1}{\gmr}\fdv{G}{P}\,\nabla P}
-\fdv{F}{\lr{\gmr}}\nabla\cdot\fdv{G}{\bol{u}}-\frac{1}{\gmr}\fdv{F}{P}\fdv{G}{\bol{u}}\cdot\nabla P} \notag\\
&=\int_\Omega\dd^3 x\,\lrs{\frac{\bol{\omega}}{\gmr}\cdot\fdv{F}{\bol{u}}\cp\fdv{G}{\bol{u}}
+\fdv{G}{\bol{u}}\cdot\nabla\fdv{F}{\lr{\gmr}}-\fdv{F}{\bol{u}}\cdot\nabla\fdv{G}{\lr{\gmr}}
+\frac{\nabla P}{\gmr}\cdot\lr{\fdv{G}{P}\fdv{F}{\bol{u}}-\fdv{F}{P}\fdv{G}{\bol{u}}}
}. \label{PB}
\end{align}
This Poisson bracket satisfies all Poisson bracket axioms, including antisymmetry and the Jacobi identity (see Appendix \ref{sub: Proof that Poisson Bracket (Poisson-bracket-gen3D) Satisfies Jacobi Identity}), under the boundary condition \eqref{gen3D-functionals-boundary}.

Using the Poisson bracket (\ref{PB}), we can identify the necessary and sufficient conditions for a functional $C\lrs{\bol{u},\gmr,P}$ to be a Casimir invariant, 
\begin{subequations} \label{Casimir-gen3D}
\begin{align}
\bol{\omega}\cp\fdv{C}{\bol{u}}+\gmr\,\nabla\fdv{C}{\lr{\gmr}}-\fdv{C}{P}\,\nabla P &=\bol{0} \label{Casimir-gen3D-1},\\
\nabla\cdot\frac{\delta C}{\delta\bol{u}} &=0 ,\\
\fdv{C}{\bol{u}}\cdot\nabla P &=0 \label{Casimir-gen3D-3}.
\end{align}
\end{subequations}
For instance, the total mass of the system 
\eq{
M=\int_{\Omega}\gmr\,\dd^3x,\label{M}
}
satisfies the system \eqref{Casimir-gen3D}, and therefore qualifies as 
a Casimir invariant. Its constancy holds independently of the specific form of $\mc{H}$, as 
\eq{
\frac{\dd M}{\dd t}=\lrc{M,\mc{H}}=0~~~~\forall~\mc{H}.
}
We note, however, that the relativistic helicity defined later \eqref{hel} is not a Casimir invariant of the Poisson bracket \eqref{PB} because it does not satisfy the condition \eqref{Casimir-gen3D-3}. 

Finally, we remark that the same results hold if periodic boundary conditions are imposed instead of \eqref{un} and (\ref{gen3D-functionals-boundary}).

\section{A Relativistic Equation of State}
The purpose of this section is to introduce the concept of a $\gamma$-barotropic equation of state, which relates the pressure \( P \) to the mass density \( \varrho_0 \) and the Lorentz factor \( \gamma \). This equation of state will be employed in subsequent sections to derive a reduced form of the relativistic Euler equations that conserves helicity in 3D and enstrophy in 2D.

We refer to any equation of state of the form
\begin{equation}
P = P(\gamma \varrho_0) \label{gbar}
\end{equation}
as \textit{\(\gamma\)-barotropic}. The specific case \( P \propto \gamma \varrho_0 \) will be called the \textit{\(\gamma\)-ideal equation of state}.

In the non-relativistic limit \( c \rightarrow +\infty \), the $\gamma$-barotropic and $\gamma$-ideal equations of state naturally reduce to the barotropic and ideal equations of state, respectively.

Mathematically, the $\gamma$-barotropic equation of state \eqref{gbar} allows us to reduce the relativistic Euler equations \eqref{EoM4} to two evolution equations governing the mass density $\varrho_0$ and the velocity field $\bol{u}$. Additionally, it results in the vanishing of the relativistic analog of the classical baroclinic effect in the vorticity equation, thereby ensuring the conservation of relativistic helicity and enstrophy (see Section 7 for further details).

It is beyond the scope of this study to derive the $\gamma$-barotropic equation of state \eqref{gbar} from first principles, as this requires a consistent formulation of thermodynamics and statistical mechanics within a relativistic framework—a problem that remains open. In particular, there is no consensus on how thermodynamic quantities such as temperature, heat, and pressure transform under Lorentz transformations (see, e.g., \cite{Heras,Farias,Tol}). This ambiguity arises because the Lorentz transformation of thermodynamic variables depends on specific assumptions about the form of thermodynamic laws in a relativistic context, especially regarding the covariance of the first and second laws of thermodynamics \cite{Mares}.

Some additional remarks on the $\gamma$-barotropic equation of state \eqref{gbar} are in order. First, due to the presence of the Lorentz factor $\gamma$, this equation of state is not invariant under Lorentz transformations. This is not necessarily problematic, as thermodynamic quantities are not generally considered covariant observables. Moreover, the pressure \( P \) appearing in the energy-momentum tensor of a perfect fluid assumes a balance between gravitational force and hydrostatic force due to pressure in a weak field regime. This pressure is thus associated with a specific notion of steady equilibrium in a coordinate system where such a force balance is realized, justifying the dependence of \( P \) on \( \gamma \).

Secondly, we emphasize that the $\gamma$-barotropic equation of state \eqref{gbar} does not describe the Lorentz transformation of pressure, meaning it does not relate the pressure \( P \) observed in one inertial frame \( I \) to the pressure \( P' \) in another inertial frame \( I' \). Rather, it specifies a functional relationship between the fields \( P \) and \( \gamma \varrho_0 \) within the frame \( I \). This functional relationship is mathematically compatible with the system \eqref{EoM4} and does not require determining the Lorentz transformation of pressure.

In this work, we adopt the $\gamma$-barotropic equation of state \eqref{gbar} as it is mathematically consistent and free from physical contradictions. However, it remains outside the scope of this study to establish the conditions under which physical systems obeying \eqref{gbar} might actually be realized.

For further context on the definition of thermodynamic quantities in relativity, see \cite{Rovelli} for an in-depth treatment of statistical mechanics in general relativity, which introduces a corresponding notion of temperature.

\section{Perfect $\gamma$-Barotropic Fluids}

In this section, we derive the equations of motion governing the dynamics of a perfect 3D fluid subject to a $\gamma$-barotropic equation of state \eqref{gbar} and their noncanonical Hamiltonian Structure. 

\subsection{Relativistic Euler Equations for a Perfect $\gamma$-Barotropic Fluid}

We begin by introducing the velocity field 
\eq{
\boldsymbol{v} = \frac{\boldsymbol{u}}{\gamma},}
which represents the velocity measured by a static observer in coordinate time $t$. 
Observe that
\begin{equation}
\gamma^2=1+\frac{\bol{u}^2}{c^2}=\frac{1}{1-\frac{\bol{v}^2}{c^2}}.  
\end{equation}
Substituting the equation of state $P=P\lr{\gmr}$ into system \eqref{EoM4}, we obtain the governing equations 
\begin{subequations} \label{EoM5}
\begin{align}
&\nabla\cdot\bol{v}=0.\label{divv}
\\
&\frac{\p}{\p t}\lr{\gmr}=-\bol{v}\cdot\nabla\lr{\gmr},\label{con}
\\
&\frac{\p\bol{u}}{\p t} =\frac{1}{\gamma}\,\bol{u}\cp\bol{\omega}-c^2\,\nabla\gamma-\frac{1}{\gmr}\,\nabla P\lr{\gmr}.\label{mom}
\end{align}
\end{subequations} 
Equation \eqref{divv} expresses the \ti{spatial incompressibility} of the 
velocity field $\bol{v}$, while \eqref{con} and \eqref{mom} are the continuity and momentum equations. 

Let 
\eq{
\bol{\omega}=\nabla\cp\bol{u}=\nabla\cp\lr{\gamma\bol{v}},
}
denote the vorticity vector associated with the velocity field $\bol{u}$. 
From the governing equations (\ref{EoM5}), we see that 
the vorticity equation and the evolution equation for the Lorentz factor take the form 
\begin{subequations}
\begin{align}
&\frac{\p\bol{\omega}}{\p t} =\nabla\cp\lr{\frac{1}{\gamma}\,\bol{u}\cp\bol{\omega}}, \label{omega_t}\\
&\frac{\p\gamma}{\p t} =-\frac{1}{c^2\gamma}\,\bol{u}\cdot\lr{c^2\nabla\gamma+\frac{1}{\gmr}\,\nabla P} .
\end{align}
\end{subequations}


\subsection{Noncanonical Hamiltonian Structure and Casimir Invariants}
\label{sec: Hamiltonian Structure and Casimir Invariants}

Let us determine the noncanonical Hamiltonian structure
associated with the governing equations \eqref{EoM5} for a perfect fluid subject to a $\gamma$-barotropic equation of state. 

The Hamiltonian of the system can be inferred from the Hamiltonian \eqref{H1} of the parent system and it is given by
\begin{equation}
\mc{H}\lrs{\bol{u},\gmr}=\int_\Omega\lrs{c^2\gamma^2\vrho+\zeta\lr{\gmr}}\,\dd^3 x=\int_\Omega\lrs{c\gmr\,\sqrt{c^2+\bol{u}^2}+\zeta\lr{\gmr}}\,\dd^3x,\label{H2}
\end{equation}
where $\zeta$ denotes a function such that $\zeta''\lr{\gmr}=P'\lr{\gmr}/\gmr$, with $'$ denoting differentiation with respect to $\gmr$.
Using \eqref{EoM5}, it can be shown that 
\begin{align}
\frac{\dd\mc{H}}{\dd t}
&=-\int_\Omega\nabla\cdot\lrc{\lrs{c^2\gmr+\frac{P\lr{\gmr}+\zeta\lr{\gmr}}{\gamma}}\bol{u}}\dd^3x.
\end{align}
It follows that the Hamiltonian \eqref{H2} is a constant of motion under the boundary condition 
\begin{equation}
\lrs{c^2\gmr+\frac{P\lr{\gmr}+\zeta\lr{\gmr}}{\gamma}}\bol{u}\cdot\bol{n}=0~~~~{\rm on}~~\p\Omega.
\label{bc2}
\end{equation}
Next, observe that we take as independent variables the pair $\lrc{\bol{u},\gmr}$, and not $\lrc{\bol{u},\vrho}$.
Then, the functional derivatives of the Hamiltonian \eqref{H2} can be evaluated as follows: 
\begin{equation}
\frac{\delta \mc{H}}{\delta\bol{u}}=\vrho\bol{u},~~~~
\frac{\delta \mc{H}}{\delta\lr{\gmr}}=c^2\gamma+\zeta'\lr{\gmr}. 
\end{equation}
Restricting the state space of the velocity field $\bol{u}$ to those vector fields satisfying
$\nabla\cdot\bol{v}=\nabla\cdot\lr{{\bol{u}/\gamma}}=0$, the governing equations \eqref{EoM5} can thus be obtained through the Poisson operator
\begin{equation}
\begin{bmatrix}
-\frac{1}{\gmr}\bol{\omega}\cp&-\nabla\\-\nabla\cdot&0
\end{bmatrix},
\end{equation}
and the Poisson bracket 
\begin{align}
\lrc{F,G}
&=\int_\Omega\dd^3 x\,\lrs{\frac{\delta F}{\delta\bol{u}}\cdot\lr{-\frac{1}{\gmr}\,\bol{\omega}\times\frac{\delta G}{\delta\bol{u}}-\nabla\frac{\delta G}{\delta\lr{\gmr}}}-\frac{\delta F}{\delta\lr{\gmr}}\nabla\cdot\frac{\delta G}{\delta\bol{u}}} \notag\\
&=\int_\Omega\dd^3 x\,\lrs{\frac{1}{\gmr}\,\bol{\omega}\cdot\lr{\frac{\delta F}{\delta\bol{u}}\times\frac{\delta G}{\delta\bol{u}}}+\frac{\delta G}{\delta\bol{u}}\cdot\nabla\frac{\delta F}{\delta\lr{\gmr}}-\frac{\delta F}{\delta\bol{u}}\cdot\nabla\frac{\delta G}{\delta\lr{\gmr}}}, 
\label{Poisson-bracket-3D}
\end{align}
with the boundary condition
\begin{equation}
\frac{\delta F}{\delta\bol{u}}\cdot\bol{n}=0~~~~{\rm on}~~\p\Omega
\label{3D-functionals-boundary}
\end{equation}
for all functionals $F\lrs{\bol{u},\gmr}$.
Again, we emphasize that the same Hamiltonian structure can be obtained if periodic boundary conditions are imposed instead of equation (\ref{3D-functionals-boundary}).
The bracket \eqref{Poisson-bracket-3D} satisfies all Poisson bracket axioms, including antisymmetry and the Jacobi identity (see Appendix \ref{sub: Proof that Poisson Bracket (Poisson-bracket-3D) Satisfies Jacobi Identity}).

From the Poisson bracket (\ref{Poisson-bracket-3D}), we can derive the necessary and sufficient conditions for a functional $C\lrs{\bol{u},\gmr\,}$ to be a Casimir invariant. We find 
\begin{subequations} \label{Casimir-3D}
\begin{align}
\bol{\omega}\cp\frac{\delta C}{\delta\bol{u}}+\nabla\frac{\delta C}{\delta\lr{\gmr}} &=\bol{0} \label{Casimir-3D-1},\\
\nabla\cdot\frac{\delta C}{\delta\bol{u}} &=0 .
\end{align}
\end{subequations}

It is clear that the relativistic helicity 
\begin{align}
K=\int_\Omega\bol{u}\cdot\bol{\omega}\,\dd^3x
=\int_{\Omega}\gamma\bol{v}\cdot\nabla\cp\lr{\gamma\bol{v}}\,\dd^3x,
\label{hel}
\end{align}
is a Casimir invariant, provided that variations $\delta\bol{u}$ of the velocity field $\bol{u}$ satisfy the boundary condition
\begin{equation}
\bol{n}\cp\delta\bol{u}=0~~~~{\rm on}~~\p\Omega
\label{hel-Casimir-boundary}.
\end{equation}
Indeed, since
\begin{equation}
\delta K=\int_\Omega \lrs{\delta\bol{u}\cdot2\,\bol{\omega}+\nabla\cdot\lr{\delta\bol{u}\cp\bol{u}}}\dd^3x
=\int_\Omega \delta\bol{u}\cdot2\,\bol{\omega}\,\dd^3x+\int_{\p\Omega}\bol{u}\cdot\lr{\bol{n}\cp\delta\bol{u}}\dd S ,
\end{equation}
we obtain
\begin{equation}
\frac{\delta K}{\delta\bol{u}}=2\,\bol{\omega},~~~~\frac{\delta K}{\delta\lr{\gmr}}=0
\label{hel-fdv} ,
\end{equation}
which satisfies the necessary and sufficient conditions \eqref{Casimir-3D} to be a Casimir invariant. 

Finally, observe that the total mass $M$ defined in equation  \eqref{M} remains a Casimir invariant of the reduced Poisson bracket \eqref{Poisson-bracket-3D}.

\section{2D Reduction of a Perfect $\gamma$-Barotropic Fluid}

In this section, we derive the equations of motion governing the dynamics of a perfect 2D fluid subject to a $\gamma$-barotropic equation of state \eqref{gbar} and discuss their noncanonical Hamiltonian Structure. 

\subsection{{Stream Function Formulation of 2D Perfect $\gamma$-Barotropic Fluids}}
\label{sec: 2D Relativistic Euler Equations with a Stream Function}

To explore the two-dimensional reduction of system \eqref{EoM5}, we assume the fluid exhibits symmetry in the \( x^3 \)-direction. More specifically, we take \( \partial_3 = \partial / \partial x^3 \) as a Killing field, implying a symmetry in the system such that the variables \( u^\mu \) for \( \mu = 0, \ldots, 3 \), along with \( \rho_0 \) and \( P \), are independent of \( x^3 \), and the \( x^3 \)-component of the four-velocity, \( u^3 \), vanishes everywhere.

The domain of the system is defined as \( \Omega = \Sigma \times [0,1] \), where \( \Sigma \subset \mathbb{R}^2 \) is a two-dimensional smooth bounded domain. 

From here, we restrict our analysis to the case where the velocity field is represented as 
\eq{
\boldsymbol{v} = \nabla \Psi \times \nabla z,\label{v2d}
}
with \( z = x^3 \) and the stream function \( \Psi(x^0, x^1, x^2) \). This form of the velocity field clearly satisfies the incompressibility condition \eqref{divv}.
Since
\eq{
\boldsymbol{\omega} = -\nabla \cdot (\gamma \nabla \Psi) \nabla z = \omega_z \nabla z,\label{o2d}
}
the vorticity \( \boldsymbol{\omega} \) only has a vertical \( z \)-component. Substituting equations \eqref{v2d} and \eqref{o2d} into (\ref{omega_t}), the vorticity equation can be written as a time evolution equation for the stream function  $\Psi$: 
\begin{equation}
\frac{\p}{\p t}\,\nabla\cdot\lr{\gamma\nabla\Psi}=\lrs{\Psi,\nabla\cdot\lr{\gamma\nabla\Psi}} ,\label{psit}
\end{equation}
where the square bracket on the right-hand side is defined as
\begin{equation}
\lrs{f,g}=\nabla z\cdot\nabla f\cp\nabla g 
\end{equation}
for arbitrary two-dimensional functions $f$, $g$.
Observe that, in the non-relativistic limit $c\rightarrow+\infty$, equation \eqref{psit} reduces to the evolution equation for the stream function of a classical two-dimensional fluid.  

In summary, the governing equations for two-dimensional perfect $\gamma$-barotropic flows are
\begin{subequations}
\begin{align}
\frac{\p}{\p t}\lr{\gmr} &=\lrs{\Psi,\gmr} ,
\label{2D-EoM-gmr}
\\
\frac{\p}{\p t}\,\tlaplacian\Psi &=\lrs{\Psi,\tlaplacian\Psi} ,
\label{2D-EoM-Psi}
\end{align}\label{EoM2D}
\end{subequations}
where we defined the $\gamma$-Laplacian
\eq{
\tlaplacian=\nabla\cdot\lr{\gamma\nabla}.}
Note that equation (\ref{2D-EoM-Psi}) can be solved independently, and once a solution for \( \Psi \) is obtained, it can be substituted into equation (\ref{2D-EoM-gmr}), which is then solved for \( \vrho \). Additionally, it is worth mentioning that if \( \gmr \) is a function of \( \tlaplacian \Psi \), the continuity equation (\ref{2D-EoM-gmr}) is automatically satisfied, as it coincides with equation (\ref{2D-EoM-Psi}).

\subsection{Noncanonical Hamiltonian Structure and Casimir Invariants}
Our goal in this subsection is to further simplify system \eqref{EoM2D} by eliminating the continuity equation under the assumption that $\gmr$ is a positive constant. This reduction leads to a single evolution equation for the stream function $\Psi$, whose bracket structure becomes the Poisson bracket structure of a classical two-dimensional fluid in the non-relativistic limit. This result also establishes a direct analogy with the Hamiltonian structure of the Charney equation in geophysical fluid dynamics and the Hasegawa--Mima equation in plasma physics \cite{Tassi2009,SY24}.

Given that the assumption $\gmr = k \in \mathbb{R}_{>0}$ would imply $\varrho_0 \rightarrow 0$ in the limit $\abs{\bol{v}}\rightarrow c$, we confine our analysis to the regime where $\abs{\bol{v}} \ll c$. Under these conditions, only equation (\ref{2D-EoM-Psi}) remains relevant for consideration.

For small $\bol{v}^2/c^2$, we can approximate the Hamiltonian \eqref{H2} as follows 
\begin{equation}
\mc{H}=\int_\Sigma \lrs{c^2 \gamma^2\vrho +\zeta\lr{\gmr}}\dd^2x
\approx \int_\Sigma \lrs{c^2\vrho+\zeta\lr{\gmr}}\,\dd^2x .
\end{equation}
By setting $\gmr=k\in\mathbb{R}_{>0}$, we then obtain
\begin{equation}
\mc{H}_{2D}\lrs{\Psi}=c^2k\int_{\Sigma}\gamma^{-1}\,\dd^2x
=c^2k\int_{\Sigma}\sqrt{1-\frac{1}{c^2}\abs{\nabla\Psi}^2}\,\dd^2x.
\label{H2D}
\end{equation}
Here, it is worth observing that one arrives at the same Hamiltonian even if the order of the approximations is reversed. 
Indeed, setting $\gamma\varrho_0=k$ first in $\mc{H}$, and then assuming $\bol{v}^2/c^2$ to be small leads to
\eq{
\mc{H}=\int_{\Sigma}\lrs{k\gamma+\zeta\lr{k}}\,\dd^2 x\approx \int_{\Sigma}\lrs{c^2k\lr{2-1+\frac{1}{2}\frac{\bol{v}^2}{c^2}}+\zeta\lr{k}}\,\dd^2 x\approx-\mc{H}_{2D}+\lrs{2c^2k+\zeta\lr{k}}
{{\Sigma}}.
}
Since $\gmr$ is constant, the pressure term $\zeta\lr{\gmr}$ can be neglected from the analysis. 
Using \eqref{con} and \eqref{mom},
it can be shown that
\begin{equation}
\frac{\dd\mc{H}_{2D}}{\dd t} 
=-c^2k\int_\Sigma \nabla\cdot\lr{\frac{\bol{v}}{\gamma}}\dd^2x
=-c^2k\int_{\p\Sigma} \frac{1}{\gamma}\,\nabla z\cdot\bol{n}\cp\nabla\Psi\,\dd^2x .
\end{equation}
It follows that the Hamiltonian \eqref{H2D} is a constant of motion under the boundary condition 
\begin{equation}
\Psi=\const~~~~{\rm on}~~\p\Sigma,
\end{equation}
which implies 
$\bol{u}\cdot\bol{n}=0$ on $\p\Sigma$. 

The candidate Poisson bracket for the 2D flow can now be defined as
\begin{align}
\lrc{F,G}_{2D}
&=\frac{1}{k}\int_\Sigma \tlaplacian\Psi\lrs{\,\tlaplacian^{-1}\frac{\delta F}{\delta\Psi}\,,\,\tlaplacian^{-1}\frac{\delta G}{\delta\Psi}\,}\dd^2x
\label{Poisson-bracket-2D}
\end{align}
for arbitrary 2D functionals $F\lrs{\Psi}$, $G\lrs{\Psi}$
obeying the boundary condition
\begin{equation}
\tlaplacian^{-1}\fdv{F}{\Psi}=\const~~~~{\rm on}~~\p\Sigma,
\label{2D-F-boundary}
\end{equation}
where $\Delta_{\gamma}^{-1}$ denotes the inverse $\gamma$-Laplacian. In the following, 
we will assume the existence of such inverse operator.
{%
The bracket \eqref{Poisson-bracket-2D} clearly satisfies antisymmetry and correctly reproduces the Poisson bracket for two-dimensional flow in the non-relativistic limit $c \to +\infty$ ($\gamma \to 1$). However, proving the Jacobi identity for the antisymmetric bracket \eqref{Poisson-bracket-2D} when $\gamma \neq 1$ is complicated by the functional derivatives of the inverse operator $\Delta_{\gamma}^{-1}$. As a result, the proof is not provided here and remains an open problem.
Nonetheless, as we will show, the proposed Poisson bracket \eqref{Poisson-bracket-2D} not only reproduces the classical limit correctly, but also possesses enstrophy as a Casimir invariant.
}%

Let us verify that the bracket 
\eqref{Poisson-bracket-2D} generates the equations of motion.
Using the boundary condition \eqref{2D-F-boundary} on functionals, the following identity holds:
\begin{align}
&\int_\Sigma \tlaplacian\Psi\lrs{\tlaplacian^{-1}\fdv{F}{\Psi},\tlaplacian^{-1}\fdv{G}{\Psi}}\dd^2x
+\int_\Sigma \tlaplacian^{-1}\fdv{F}{\Psi}\lrs{\tlaplacian\Psi,\tlaplacian^{-1}\fdv{G}{\Psi}}\dd^2x
\notag \\ &
=\int_{\p\Sigma}\tlaplacian^{-1}\fdv{F}{\Psi}\,\tlaplacian\Psi\lr{\nabla z\cdot\bol{n}\cp\nabla \tlaplacian^{-1}\fdv{G}{\Psi}}\dd l
=0.
\label{2D-formula}
\end{align}
Here, $\dd l$ is the line element on $\p\Sigma$. 
We will now use equation \eqref{2D-formula}  to derive the equations of motion \eqref{Hamilton-eq-2D}  and {the conditions \eqref{Casimir-2D} for a functional to be a Casimir invariant}.

When we calculate functional derivatives, we further assume that variations $\delta\Psi$ of the stream function $\Psi$ satisfy the boundary conditions 
\begin{equation}
\delta\Psi=0,~~~~\bol{n}\cdot\nabla\delta\Psi=0~~~~{\rm on}~~\p\Sigma, 
\label{boundary-deltaPsi}
\end{equation}
where \( \boldsymbol{n} \) represents the unit outward normal to the boundary \( \partial \Sigma \).
Then, noting that 
\begin{align}
\delta\mc{H}_{2D}
&=-c^2k\int_\Sigma\frac{\gamma}{c^2}\,\nabla\Psi\cdot\nabla\delta\Psi\,\dd^2x
=k\int_\Sigma\delta\Psi\,\nabla\cdot\lr{\gamma\nabla\Psi}\dd^2x
-k\int_{\p\Sigma}\delta\Psi\lr{\gamma\nabla\Psi}\cdot\bol{n}\,\dd l
, \\
\delta\lrs{\Delta_{\gamma\lr{\bol{\xi}}}\Psi\lr{\bol{\xi}}}
&=\int_\Sigma \delta^{\lr{2}}\lr{\bol{x}-\bol{\xi}}\Delta_{\gamma\lr{\bol{x}}}\delta\Psi\lr{\bol{x}}\dd^2x
\notag \\
&=\int_{\p\Sigma}\bol{n}\lr{\bol{x}}\cdot\lrs{\gamma\lr{\bol{x}}\delta^{\lr{2}}\lr{\bol{x}-\bol{\xi}}\nabla\delta\Psi\lr{\bol{x}}
-\delta\Psi\lr{\bol{x}}\gamma\lr{\bol{x}}\nabla\delta^{\lr{2}}\lr{\bol{x}-\bol{\xi}}}\dd l\lr{\bol{x}}
\notag\\ &\qquad\qquad
+\int_\Sigma \delta\Psi\lr{\bol{x}}\Delta_{\gamma\lr{\bol{x}}}\delta^{\lr{2}}\lr{\bol{x}-\bol{\xi}}\dd^2x ,
\end{align}
the functional derivatives of $\mc{H}_{2D}$ and $\tlaplacian\Psi$ can be evaluated to be 
\begin{equation}
\frac{\delta\mc{H}_{2D}}{\delta\Psi}
=k\,\tlaplacian\Psi ,~~~~
\frac{\delta}{\delta\Psi\lr{\bol{x}}}\Delta_{\gamma\lr{\bol{\xi}}}\Psi\lr{\bol{\xi}}
=\Delta_{\gamma\lr{\bol{x}}}\delta^{\lr{2}}\lr{\bol{x}-\bol{\xi}} 
\end{equation}
where $\bol{\xi}\in\Sigma$ and $\delta^{\lr{2}}$ denotes the 2D delta function. 


Thus, {using \eqref{2D-formula},} we obtain
\begin{equation}
\lrc{\tlaplacian\Psi,\mc{H}_{2D}}_{2D}
=\frac{1}{k}\int_\Sigma\tlaplacian\Psi\lrs{\delta^{\lr{2}},k\Psi}\,\dd^2x=-\lrs{\tlaplacian\Psi,\Psi}. 
\label{2D-EoM-mid}
\end{equation}
Hence, equation 
\begin{equation}
\frac{\p}{\p t}\,\tlaplacian\Psi=\lrc{\tlaplacian\Psi,\mc{H}_{2D}}_{2D}
\label{Hamilton-eq-2D}
\end{equation}
corresponds to the evolution  equation (\ref{2D-EoM-Psi}) for the stream function as desired.

Next, from equations (\ref{Poisson-bracket-3D}) and \eqref{2D-formula}, we see that the necessary and sufficient condition for a functional $C\lrs{\Psi}$ to be a Casimir invariant of the bracket \eqref{Poisson-bracket-2D} is
\begin{equation}
\lrs{\,\tlaplacian\Psi\,,\,\tlaplacian^{-1}\frac{\delta C}{\delta\Psi}\,}=0~~~~{\rm in}~~\Sigma.
\label{Casimir-2D}
\end{equation}
Noting that the relativistic enstrophy
\begin{equation}
E=\frac{1}{2}\int_\Sigma\frac{\bol{\omega}^2}{\gmr}\,\dd^2x
=\frac{1}{2}\int_\Sigma\frac{\abs{\nabla\cp\lr{\gamma\bol{v}}}^2}{\gmr}\,\dd^2x
\label{enstrophy-definition-general}
\end{equation}
reduces to the form
\begin{equation}
E=\frac{1}{2k}\int_\Sigma\lr{\tlaplacian\Psi}^2\dd^2x .
\label{enstrophy-definition-Psi}
\end{equation}
under the condition that  $\gmr$ is constant, one can show that it is a Casimir invariant of the bracket \eqref{Poisson-bracket-2D} under the boundary conditions \eqref{boundary-deltaPsi} for $\delta\Psi$.
Indeed, since
\begin{align}
\delta E
&=\frac{1}{k}\int_\Sigma\lrs{\delta\Psi\,\tlaplacian^2\Psi+\nabla\cdot\lrc{\gamma\,\tlaplacian\Psi\,\nabla\delta\Psi-\gamma\,\delta\Psi\,\nabla\lr{\tlaplacian\Psi}}}\dd^2x
\notag \\ &
=\frac{1}{k}\int_\Sigma\delta\Psi\,\tlaplacian^2\Psi~\dd^2x
+\frac{1}{k}\int_{\p\Sigma}\lrs{\gamma\,\tlaplacian\Psi\lr{\bol{n}\cdot\nabla\delta\Psi}-\gamma\,\delta\Psi\lr{\bol{n}\cdot\nabla\lr{\tlaplacian\Psi}}}\dd l
\end{align}
and the boundary terms vanish, we obtain
\begin{equation}
\frac{\delta E}{\delta\Psi}=\frac{1}{k}\,\tlaplacian^2\Psi \label{ens-fdv} ,
\end{equation}
which satisfies the necessary and sufficient condition  \eqref{Casimir-2D} to be a Casimir invariant. 
%



\section{Breakdown of Helicity and Enstrophy Conservation}
In this section, we examine the breakdown of helicity and enstrophy conservation in the parent system \eqref{EoM4}, which is not constrained by a \(\gamma\)-barotropic equation of state. Specifically, we will show that the presence of the \(\gamma\)-\textit{baroclinic effect}:
\eq{
\bol{b}_{\gamma}=-\nabla\left(\frac{1}{\gmr}\right) \times \nabla P,
}
appearing in the vorticity equation
\eq{
\frac{\p\bol{\omega}}{\p t}=\nabla\cp\lr{\frac{1}{\gamma}\bol{u}\cp\bol{\omega}}-\nabla\left(\frac{1}{\gmr}\right) \times \nabla P, \label{voreq-b}
}
leads to the violation of both helicity and enstrophy conservation in such regimes.

\subsection{Breakdown of Helicity Conservation}

From the equations of motion \eqref{EoM4}, we have
\begin{align}
&\frac{\p\bol{u}}{\p t}\cdot\bol{\omega} =\nabla\cdot\lr{\bol{u}\cp\frac{\p\bol{u}}{\p t}}+\bol{u}\cdot{\frac{\p\bol{\omega}}{\p t}} ,\\
&\bol{u}\cdot\frac{\p\bol{\omega}}{\p t} 
=\bol{u}\cdot\nabla\cp\lr{\frac{1}{\gamma}\,\bol{u}\cp\bol{\omega}}+\bol{u}\cdot\bol{b}_{\gamma} 
=\nabla\cdot\lrs{\frac{1}{\gamma}\lr{\bol{u}\cp\bol{\omega}}\cp\bol{u}}+\bol{u}\cdot\bol{b}_{\gamma},
\end{align}
It follows that the time derivative of the helicity $K$ is: 
\begin{align}
\frac{\dd K}{\dd t}
&=\int_\Omega\lr{\bol{u}\cdot\frac{\p\bol{\omega}}{\p t}+\bol{\omega}\cdot\frac{\p\bol{u}}{\p t}}\dd^3x
\notag \\ &
=\int_\Omega\nabla\cdot\lrs{\bol{u}\cp\frac{\p\bol{u}}{\p  t}+\frac{2}{\gamma}\lr{\bol{u}\cp\bol{\omega}}\cp\bol{u}}\dd^3x+{2}\int_\Omega 
\bol{u}\cdot\bol{b}_{\gamma}\,\dd^3x.\label{dKdt}
\end{align}
The divergence term vanishes under the boundary conditions 
\begin{equation}
\bol{u}\cdot\bol{n}=0,~~~~\bol{\omega}\cdot\bol{n}=0,~~~~\frac{\p\bol{u}}{\p t}\cp\bol{n}={\bol{0}}~~~~
{\rm on}~~\p\Omega.\label{bcK}
\end{equation}
Here, it is worth observing that these boundary conditions 
represent {special cases of the boundary conditions \eqref{bc2}, \eqref{3D-functionals-boundary}, and \eqref{hel-Casimir-boundary}}. 
However, the second term in equation \eqref{dKdt} involving the $\gamma$-baroclinic effect $\bol{b}_{\gamma}$ does not vanish in general.
Thus, the relativistic helicity \eqref{hel} is not an invariant of system \eqref{EoM4}.
This fact is consistent with the earlier observation that 
$K$ is not a Casimir invariant of the Poisson bracket \eqref{PB} because 
\begin{equation}
\fdv{K}{\bol{u}}\cdot\nabla P=2\,\bol{\omega}\cdot\nabla P\neq0, 
\end{equation}
which conflicts with the Casimir invariant condition  \eqref{Casimir-gen3D-3}.

Finally, observe that $\bol{b}_{\gamma}$ identically vanishes under a $\gamma$-barotropic equation of state, implying conservation of helicity under the boundary conditions \eqref{bcK} as discussed in section \ref{sec: Hamiltonian Structure and Casimir Invariants}.



\subsection{Breakdown of Enstrophy Conservation}

Let us move on to 2D perfect $\gamma$-barotropic flows (which do not necessarily have constant $\gmr$ as described in section \ref{sec: 2D Relativistic Euler Equations with a Stream Function}).
From the vorticity equation \eqref{voreq-b}, we have
\begin{gather}
\bol{v}\cp\bol{\omega}=-\omega_z \nabla\Psi ,~~~~
\bol{\omega}_t=-\nabla\omega_z \cp\nabla\Psi+{\bol{b}_{\gamma}} ,~~~~
\bol{\omega}\cdot\bol{\omega}_t=-\frac{1}{2}\,\bol{v}\cdot\nabla\omega_z^2+{\bol{\omega}\cdot\bol{b}_{\gamma}}
, \\
\frac{\bol{\omega}^2}{2\lr{\gmr}^2}\,\bol{v}\cdot\nabla\lr{\gmr}
=-\frac{1}{2}\,\omega_z^2\,\bol{v}\cdot\nabla\lr{\frac{1}{\gmr}}
=-\bol{v}\cdot\nabla\lr{\frac{\omega_z^2}{2\gmr}}+\frac{1}{2\gmr}\,\bol{v}\cdot\nabla\omega_z^2
,
\end{gather}
It follows that the time derivative of the enstrophy $E$ is
\begin{equation}
\frac{\dd E}{\dd t}
=\int_\Sigma\lrs{\frac{\bol{\omega}\cdot\bol{\omega}_t}{\gmr}+\frac{\bol{\omega}^2}{2\lr{\gmr}^2}\,\bol{v}\cdot\nabla\lr{\gmr}}\dd^2x
=\int_\Sigma \lrs{
-\nabla\cdot\lr{\frac{\omega_z^2}{2\gmr}\,\bol{v}}
+\frac{\omega_z}{\gmr}\,\nabla z\cdot\bol{b}_\gamma
} \dd^2x.\label{dEdt}
\end{equation}
The divergence term vanishes under the boundary conditions 
\begin{equation}
\Psi=\const~~~~{\rm on}~~\p\Sigma,
\end{equation}
which corresponds to $\bol{u}\cdot\bol{n}=0$ on $\p\Sigma$.
However, the second term in equation \eqref{dEdt} involving the $\gamma$-baroclinic effect $\bol{b}_{\gamma}$ does not vanish in general.
Thus, the relativistic enstrophy \eqref{enstrophy-definition-general} is not a constant of motion in the presence of $\bol{b}_{\gamma}$.
Finally, observe that $\bol{b}_{\gamma}$ identically vanishes under a $\gamma$-barotropic equation of state, leading to enstrophy conservation in $\gamma$-barotropic flows.

\section{Concluding Remarks}

In this work, we have explored the Hamiltonian structure of relativistic fluid flows and demonstrated the conservation of helicity and enstrophy in three- and two-dimensional settings, respectively, under a \(\gamma\)-barotropic equation of state. These findings provide affirmative answers to two critical questions: first, whether helicity and enstrophy can be defined in terms of three-dimensional velocity fields associated with spatial coordinates and evolving in coordinate time $t$, and second, whether these quantities correspond to Casimir invariants within the noncanonical Hamiltonian framework.

Our results confirm that relativistic helicity and enstrophy, expressed through the spatial velocity and vorticity fields, remain conserved with respect to the classical time variable. Furthermore, these invariants naturally reduce to their classical counterparts as the Lorentz factor \(\gamma\) approaches 1. This extends the well-established conservation laws of classical fluid dynamics to the relativistic regime.

The framework developed in this study opens new avenues for applying Hamiltonian theory to the modeling of astrophysical fluids and relativistic plasmas, where dissipative effects and relaxation dynamics must be carefully accounted for. The formulation of diffusion and collision operators, constrained by Casimir invariants such as helicity and enstrophy, offers a promising approach to modeling complex fluid systems in space environments. Future work will further investigate the interplay between these invariants and the physical processes governing the behavior of relativistic fluids, including the impact of baroclinic effects when the \(\gamma\)-barotropic condition is relaxed.

In summary, this study lays a solid theoretical foundation for understanding the role of helicity and enstrophy in relativistic fluid dynamics and paves the way for future applications in the study of astrophysical phenomena and relativistic plasma systems.

\section*{Acknowledgment}
T.K. and N.S. would like to acknowledge useful discussion with Z. Yoshida and H. Sugama. 

\section*{Statements and declarations}

\subsection*{Data availability}
Data sharing not applicable to this article as no datasets were generated or analysed during the current study.

\subsection*{Funding}
The research of T.K. was partially supported by JSPS KAKENHI Grant No. 23KJ0435.
The research of N.S. was partially supported by JSPS KAKENHI Grant No. 22H04936 and 24K00615. This work was partly supported by MEXT Promotion of Distinctive Joint Research Center Program JPMXP0723833165. 

\subsection*{Competing interests} 
The authors have no competing interests to declare that are relevant to the content of this article.

\appendix
\section{Intermediate Calculations}
\subsection{Derivation of equation (\ref{EoM2-0})}
\label{Derivation of Equation (EoM2-0)}

In this subsection, we derive (\ref{EoM2-0}) from (\ref{EoM1-0}).
Applying the chain rule to the left-hand side of equation (\ref{EoM1-0}), we obtain
\begin{subequations}
\begin{align}
\p_{\mu}\lr{\vrho\,u^{\mu}u^0}
&= u^0 \p_{\mu}\lr{\vrho\,u^{\mu}} + \vrho\,u^{\mu}\p_{\mu} u^0 \\
&= u^0 \p_{\mu}\lr{\vrho\,u^{\mu}} + \sum_i \frac{u^i}{u^0} \vrho\,u^{\mu}\p_{\mu} u^i \\
&= u^0 \p_{\mu}\lr{\vrho\,u^{\mu}} + \sum_i \frac{u^i}{u^0} \lrs{-\p_i P-u^i \p_{\mu}\lr{\vrho\,u^{\mu}}} \\
&= \frac{\lr{u^0}^2-\lr{u^1}^2-\lr{u^2}^2-\lr{u^3}^2}{u^0}\,\p_{\mu}\lr{\vrho\,u^{\mu}} - \sum_i \frac{1}{u^0}\,u^i \p_i P \\
&= \frac{c^2}{u^0}\,\p_{\mu}\lr{\vrho\,u^{\mu}} - \sum_i \frac{1}{u^0}\,u^i \p_i P \\
&= \frac{c^2}{u^0} \lrs{ \p_0 \lr{\vrho \sqrt{c^2+\bol{u}^2}} + \nabla\cdot\lr{\vrho\,\bol{u}} } - \frac{1}{u^0}\,\bol{u}\cdot\nabla P
\end{align}
\end{subequations}
We thus recover equation (\ref{EoM2-0}),  
\begin{equation}
\frac{c^3}{u^0}\lrs{\frac{\p}{\p t}\lr{\vrho\sqrt{1+\frac{\bol{u}^2}{c^2}}}+\nabla\cdot\lr{\vrho\,\bol{u}}}=\frac{\p P}{\p t}+\frac{\bol{u}\cdot\nabla P}{\sqrt{1+\frac{\bol{u}^2}{c^2}}} . \tagref{EoM2-0}
\end{equation}

\subsection{Proof that the Poisson bracket (\ref{PB}) satisfies the Jacobi identity}
\label{sub: Proof that Poisson Bracket (Poisson-bracket-gen3D) Satisfies Jacobi Identity}

In this subsection, we demonstrate that the Poisson bracket (\ref{PB}) satisfies the Jacobi identity.

We start by observing that the divergence terms appearing in the following expressions for the variations $\delta\bol{u}$ and  $\delta P$  
\begin{align}
\lr{\nabla\cp\delta\bol{u}}\cdot\frac{1}{\gmr}
\fdv{F}{\bol{u}}\times\fdv{G}{\bol{u}}
&=\nabla\cdot\lrs{\frac{1}{\gmr}\,\delta\bol{u}\cp\lr{\fdv{F}{\bol{u}}\cp\fdv{G}{\bol{u}}}}
+\delta\bol{u}\cdot\nabla\cp\lr{\frac{1}{\gmr}\fdv{F}{\bol{u}}\cp\fdv{G}{\bol{u}}}
\label{curl-delta-u}, \\
\frac{1}{\gmr}\,\nabla\delta P\cdot\lr{\fdv{G}{P}\fdv{F}{\bol{u}}-\fdv{F}{P}\fdv{G}{\bol{u}}}
&=\nabla\cdot\lrs{\frac{\delta P}{\gmr}\lr{\fdv{G}{P}\fdv{F}{\bol{u}}-\fdv{F}{P}\fdv{G}{\bol{u}}}}
-\delta P\,\,\nabla\cdot\lrs{\frac{1}{\gmr}\lr{\fdv{G}{P}\fdv{F}{\bol{u}}-\fdv{F}{P}\fdv{G}{\bol{u}}}}
\label{grad-delta-P}, 
\end{align}
vanish thanks to the boundary conditions \eqref{gen3D-functionals-boundary}. Indeed, the divergence term in equation \eqref{grad-delta-P} vanishes due to the boundary conditions  $\fdv{F}{\bol{u}}\cdot\bol{n}=\fdv{G}{\bol{u}}\cdot\bol{n}=0$, while the divergence term in equation \eqref{curl-delta-u} vanishes because $\fdv{F}{\bol{u}}\cp\fdv{G}{\bol{u}}$ is parallel to $\bol{n}$, which is a consequence of the fact that both $\fdv{F}{\bol{u}}$ and $\fdv{G}{\bol{u}}$ are perpendicular to $\bol{n}$.

Thus, the functional derivatives of $\lrc{F,G}$ are 
\begin{align}
\frac{\delta}{\delta u_i}\lrc{F,G}
&=\lrs{\nabla\cp\lr{\frac{1}{\gmr}\fdv{F}{\bol{u}}\cp\fdv{G}{\bol{u}}}}_i
+\frac{1}{\gmr}\,\bol{\omega}\cdot
\lr{\fdvdv{F}{u_i}{\bol{u}}\cp\fdv{G}{\bol{u}} + \fdv{F}{\bol{u}}\cp\fdvdv{G}{u_i}{\bol{u}}}
\notag\\ &\qquad
+\fdvdv{G}{u_i}{\bol{u}}\cdot\nabla\fdv{F}{\lr{\gmr}}
-\fdvdv{F}{u_i}{\lr{\gmr}}\,\nabla\cdot\fdv{G}{\bol{u}}
-\fdvdv{F}{u_i}{\bol{u}}\cdot\nabla\fdv{G}{\lr{\gmr}}
+\fdvdv{G}{u_i}{\lr{\gmr}}\,\nabla\cdot\fdv{F}{\bol{u}}
\notag\\ &\qquad
+\frac{1}{\gmr}\,\nabla P\cdot\lr{
\fdvdv{G}{u_i}{P}\fdv{F}{\bol{u}}+\fdv{G}{P}\fdvdv{F}{u_i}{\bol{u}}
-\fdvdv{F}{u_i}{P}\fdv{G}{\bol{u}}-\fdv{F}{P}\fdvdv{G}{u_i}{\bol{u}}
} ,
\end{align}
\begin{align}
\frac{\delta}{\delta\lr{\gmr}}\lrc{F,G}
&=-\frac{1}{\lr{\gmr}^2}\,\bol{\omega}\cdot\fdv{F}{\bol{u}}\cp\fdv{G}{\bol{u}}
+\frac{1}{\gmr}\,\bol{\omega}\cdot\fdvdv{F}{\bol{u}}{\lr{\gmr}}\cp\fdv{G}{\bol{u}}
+\frac{1}{\gmr}\,\bol{\omega}\cdot\fdv{F}{\bol{u}}\cp\fdvdv{G}{\bol{u}}{\lr{\gmr}}
\notag\\ &\qquad
+\fdvdv{G}{\bol{u}}{\lr{\gmr}}\cdot\nabla\fdv{F}{\lr{\gmr}}
-\fdvdv{F}{\lr{\gmr}}{\lr{\gmr}}\,\nabla\cdot\fdv{G}{\bol{u}}
\notag\\ &\qquad
-\fdvdv{F}{\bol{u}}{\lr{\gmr}}\cdot\nabla\fdv{G}{\lr{\gmr}}
+\fdvdv{G}{\lr{\gmr}}{\lr{\gmr}}\,\nabla\cdot\fdv{F}{\bol{u}}
\notag\\ &\qquad
-\frac{1}{\lr{\gmr}^2}\,\nabla P\cdot\lr{\fdv{G}{P}\fdv{F}{\bol{u}}-\fdv{F}{P}\fdv{G}{\bol{u}}}
\notag\\ &\qquad
+\frac{1}{\gmr}\,\nabla P\cdot\lr{
\fdvdv{G}{\lr{\gmr}}{P}\fdv{F}{\bol{u}}+\fdv{G}{P}\fdvdv{F}{\bol{u}}{\lr{\gmr}}
-\fdvdv{F}{\lr{\gmr}}{P}\fdv{G}{\bol{u}}-\fdv{F}{P}\fdvdv{G}{\bol{u}}{\lr{\gmr}}
} ,
\end{align}
\begin{align}
\frac{\delta}{\delta P}\lrc{F,G}
&=-\nabla\cdot\lrs{\frac{1}{\gmr}\lr{\fdv{G}{P}\fdv{F}{\bol{u}}-\fdv{F}{P}\fdv{G}{\bol{u}}}}
+\frac{\nabla P}{\gmr}\cdot\lr{\fdvdv{G}{P}{P}\fdv{F}{\bol{u}}+\fdv{G}{P}\fdvdv{F}{\bol{u}}{P}-\fdvdv{F}{P}{P}\fdv{G}{\bol{u}}-\fdv{F}{P}\fdvdv{G}{\bol{u}}{P}}
\notag\\ &\qquad
+\frac{\bol{\omega}}{\gmr}\cdot\lr{\fdvdv{F}{\bol{u}}{P}\cp\fdv{G}{\bol{u}}+\fdv{F}{\bol{u}}\cp\fdvdv{G}{\bol{u}}{P}}
\notag\\ &\qquad
+\fdvdv{G}{\bol{u}}{P}\cdot\nabla\fdv{F}{\lr{\gmr}}-\fdvdv{F}{\lr{\gmr}}{P}\,\nabla\cdot\fdv{G}{\bol{u}}
-\fdvdv{F}{\bol{u}}{P}\cdot\nabla\fdv{G}{\lr{\gmr}}+\fdvdv{G}{\lr{\gmr}}{P}\,\nabla\cdot\fdv{F}{\bol{u}} .
\end{align}
There are $11\cdot3+13+12=58$ terms in $\lrc{\lrc{F,G},H}$.
We categorize these terms by the number of functional derivatives $\delta/\delta P$, $\delta/\delta\lr{\gmr}$, and $\delta/\delta\bol{u}$ and determine whether they cancel each other out.

There are 8 terms involving two $\delta/\delta P$ and zero $\delta/\delta\lr{\gmr}$, which cancel out when cyclically summed over $F$, $G$ and $H$ as follows:
\begin{align}
&\sumcyc \frac{1}{\lr{\gmr}^2}\fdv{H}{P}\,\p_i P\,\,\nabla P\cdot
\lr{\fdv{G}{P}\fdvdv{F}{u_i}{\bol{u}}-\fdv{F}{P}\fdvdv{G}{u_i}{\bol{u}}}=0
, \\
&\sumcyc -\frac{1}{\lr{\gmr}^2}\lr{\fdv{H}{\bol{u}}\cdot\nabla P}\nabla P\cdot
\lr{\fdvdv{G}{P}{P}\fdv{F}{\bol{u}}-\fdvdv{F}{P}{P}\fdv{G}{\bol{u}}}=0
, \\
&\sumcyc \frac{1}{\lr{\gmr}^2}\fdv{H}{P}\,\p_i P\,\,\nabla P\cdot
\fdvdv{G}{u_i}{P}\fdv{F}{\bol{u}}
-\frac{1}{\lr{\gmr}^2}\lr{\fdv{H}{\bol{u}}\cdot\nabla P}\nabla P\cdot
\fdv{G}{P}\fdvdv{F}{\bol{u}}{P}=0
, \\
&\sumcyc -\frac{1}{\lr{\gmr}^2}\fdv{H}{P}\,\p_i P\,\,\nabla P\cdot
\fdvdv{F}{u_i}{P}\fdv{G}{\bol{u}}
+\frac{1}{\lr{\gmr}^2}\lr{\fdv{H}{\bol{u}}\cdot\nabla P}\nabla P\cdot
\fdv{F}{P}\fdvdv{G}{\bol{u}}{P}=0 .
\end{align}

There are 16 terms involving one $\delta/\delta P$ and one $\delta/\delta\lr{\gmr}$, which cancel out when cyclically summed over $F$, $G$ and $H$ as follows:
\begin{align}
&\sumcyc -\frac{1}{\gmr}\lr{\nabla\cdot\fdv{H}{\bol{u}}}
\fdvdv{G}{\lr{\gmr}}{P}\,\fdv{F}{\bol{u}}\cdot\nabla P
+\frac{1}{\gmr}\lr{\fdv{H}{\bol{u}}\cdot\nabla P}
\fdvdv{F}{\lr{\gmr}}{P}\,\nabla\cdot\fdv{G}{\bol{u}}=0
, \\
&\sumcyc \frac{1}{\gmr}\lr{\nabla\cdot\fdv{H}{\bol{u}}}
\fdvdv{F}{\lr{\gmr}}{P}\,\fdv{G}{\bol{u}}\cdot\nabla P
-\frac{1}{\gmr}\lr{\fdv{H}{\bol{u}}\cdot\nabla P}
\fdvdv{G}{\lr{\gmr}}{P}\,\nabla\cdot\fdv{F}{\bol{u}}=0
, \\
&\sumcyc -\frac{1}{\gmr}\,\fdv{G}{P}\lr{\fdvdv{F}{u_i}{\bol{u}}\cdot\nabla P}\p_i\fdv{H}{\lr{\gmr}}
+\frac{1}{\gmr}\fdv{H}{P}\lr{\fdvdv{G}{u_i}{\bol{u}}\cdot\nabla\fdv{F}{\lr{\gmr}}}\p_i P=0
, \\
&\sumcyc \frac{1}{\gmr}\,\fdv{F}{P}\lr{\fdvdv{G}{u_i}{\bol{u}}\cdot\nabla P}\p_i\fdv{H}{\lr{\gmr}}
-\frac{1}{\gmr}\fdv{H}{P}\lr{\fdvdv{F}{u_i}{\bol{u}}\cdot\nabla\fdv{G}{\lr{\gmr}}}\p_i P=0
, \\
&\sumcyc -\frac{1}{\gmr}\lr{\fdv{F}{\bol{u}}\cdot\nabla P}\fdvdv{G}{\bol{u}}{P}\cdot\nabla\fdv{H}{\lr{\gmr}}
+\frac{1}{\gmr}\lr{\fdv{H}{\bol{u}}\cdot\nabla P}\fdvdv{F}{\bol{u}}{P}\cdot\nabla\fdv{G}{\lr{\gmr}}=0
, \\
&\sumcyc \frac{1}{\gmr}\lr{\fdv{G}{\bol{u}}\cdot\nabla P}\fdvdv{F}{\bol{u}}{P}\cdot\nabla\fdv{H}{\lr{\gmr}}
-\frac{1}{\gmr}\lr{\fdv{H}{\bol{u}}\cdot\nabla P}\fdvdv{G}{\bol{u}}{P}\cdot\nabla\fdv{F}{\lr{\gmr}}=0
, \\
&\sumcyc -\frac{1}{\gmr}\fdv{H}{P}\lr{\fdvdv{F}{\bol{u}}{\lr{\gmr}}\cdot\nabla P}\lr{\nabla\cdot\fdv{G}{\bol{u}}}
+\frac{1}{\gmr}\lr{\nabla\cdot\fdv{H}{\bol{u}}}\fdv{F}{P}\fdvdv{G}{\bol{u}}{\lr{\gmr}}\cdot\nabla P=0
, \\
&\sumcyc \frac{1}{\gmr}\fdv{H}{P}\lr{\fdvdv{G}{\bol{u}}{\lr{\gmr}}\cdot\nabla P}\lr{\nabla\cdot\fdv{F}{\bol{u}}}
-\frac{1}{\gmr}\lr{\nabla\cdot\fdv{H}{\bol{u}}}\fdv{G}{P}\fdvdv{F}{\bol{u}}{\lr{\gmr}}\cdot\nabla P=0
.
\end{align}

There are 8 terms involving one $\delta/\delta P$, zero $\delta/\delta\lr{\gmr}$, and three   $\delta/\delta\bol{u}$, which cancel out when cyclically summed over $F$, $G$ and $H$ as follows:
\begin{align}
&\sumcyc \frac{1}{\lr{\gmr}^2}\,\epsilon_{ijk}\,\omega_i\,\fdvdv{G}{u_j}{P}\fdv{H}{u_k}\lr{\fdv{F}{\bol{u}}\cdot\nabla P}
-\frac{1}{\lr{\gmr}^2}\lr{\fdv{H}{\bol{u}}\cdot\nabla P}\bol{\omega}\cdot\fdvdv{F}{\bol{u}}{P}\cp\fdv{G}{\bol{u}}=0
, \\
&\sumcyc -\frac{1}{\lr{\gmr}^2}\,\epsilon_{ijk}\,\omega_i\,\fdvdv{F}{u_j}{P}\fdv{H}{u_k}\lr{\fdv{G}{\bol{u}}\cdot\nabla P}
-\frac{1}{\lr{\gmr}^2}\lr{\fdv{H}{\bol{u}}\cdot\nabla P}\bol{\omega}\cdot\fdv{F}{\bol{u}}\cp\fdvdv{G}{\bol{u}}{P}=0
, \\
&\sumcyc \frac{1}{\lr{\gmr}^2}\,\epsilon_{ijk}\,\omega_i\,\fdv{G}{P}\lr{\fdvdv{F}{u_j}{\bol{u}}\cdot\nabla P}\fdv{H}{u_k}
+\frac{1}{\lr{\gmr}^2}\fdv{H}{P}\lr{\bol{\omega}\cdot\fdv{F}{\bol{u}}\cp\fdvdv{G}{u_i}{\bol{u}}}\p_i P=0
, \\
&\sumcyc -\frac{1}{\lr{\gmr}^2}\,\epsilon_{ijk}\,\omega_i\,\fdv{F}{P}\lr{\fdvdv{G}{u_j}{\bol{u}}\cdot\nabla P}\fdv{H}{u_k}
+\frac{1}{\lr{\gmr}^2}\fdv{H}{P}\lr{\bol{\omega}\cdot\fdvdv{F}{u_i}{\bol{u}}\cp\fdv{G}{\bol{u}}}\p_i P=0 ,
\end{align}
with $\epsilon_{ijk}$ the Levi-Civita symbol.

There are 5 terms involving one $\delta/\delta P$, zero $\delta/\delta\lr{\gmr}$, and two $\delta/\delta\bol{u}$, which cancel out when cyclically summed over $F$, $G$ and $H$ as follows:
\begin{align}
&\sumcyc \frac{1}{\gmr}\fdv{H}{P}\,\nabla P\cdot\nabla\cp\lr{\frac{1}{\gmr}\,\fdv{F}{\bol{u}}\cp\fdv{G}{\bol{u}}}
+\frac{1}{\lr{\gmr}^2}\lr{\nabla\cdot\fdv{H}{\bol{u}}}\nabla P\cdot\lr{\fdv{G}{P}\fdv{F}{\bol{u}}-\fdv{F}{P}\fdv{G}{\bol{u}}}
\notag \\ &\qquad\qquad\qquad
+\frac{1}{\gmr}\lr{\fdv{H}{\bol{u}}\cdot\nabla P}\nabla\cdot\lrs{\frac{1}{\gmr}\lr{\fdv{G}{P}\fdv{F}{\bol{u}}-\fdv{F}{P}\fdv{G}{\bol{u}}}}
\notag \\
&=\sumcyc \frac{1}{\gmr}\fdv{H}{P}\,
\nabla\cdot\lrs{\lr{\frac{1}{\gmr}\,\fdv{F}{\bol{u}}\cp\fdv{G}{\bol{u}}}\cp\nabla P}
\notag \\ &\qquad\qquad\qquad
+\frac{1}{\lr{\gmr}^2}\fdv{H}{P}\lrc{\lr{\nabla\cdot\fdv{F}{\bol{u}}}\fdv{G}{\bol{u}}\cdot\nabla P-\lr{\nabla\cdot\fdv{G}{\bol{u}}}\fdv{F}{\bol{u}}\cdot\nabla P}
\notag \\ &\qquad\qquad\qquad
+\frac{1}{\gmr}\lr{\fdv{F}{\bol{u}}\cdot\nabla P}\nabla\cdot\lr{\frac{1}{\gmr}\fdv{H}{P}\fdv{G}{\bol{u}}}
-\frac{1}{\gmr}\lr{\fdv{G}{\bol{u}}\cdot\nabla P}\nabla\cdot\lr{\frac{1}{\gmr}\fdv{H}{P}\fdv{F}{\bol{u}}}
\notag \\
&=\sumcyc \nabla\cdot\lrs{
\frac{1}{\lr{\gmr}^2}\fdv{H}{P}\lr{\fdv{F}{\bol{u}}\cdot\nabla P}\fdv{G}{\bol{u}}
-\frac{1}{\lr{\gmr}^2}\fdv{H}{P}\lr{\fdv{G}{\bol{u}}\cdot\nabla P}\fdv{F}{\bol{u}}}
\end{align}
These terms, which appear in divergence form, must be integrated over the domain $\Omega$.  
Therefore, they vanish by application of the divergence theorem under the boundary condition (\ref{gen3D-functionals-boundary}). 
For example, 
\begin{equation}
\int_\Omega \nabla\cdot\lrs{\frac{1}{\lr{\gmr}^2}\fdv{H}{P}\lr{\fdv{F}{\bol{u}}\cdot\nabla P}\fdv{G}{\bol{u}}}\dd^3x
=\int_{\p\Omega} \frac{1}{\lr{\gmr}^2}\fdv{H}{P}\lr{\fdv{F}{\bol{u}}\cdot\nabla P}\fdv{G}{\bol{u}}\cdot\bol{n}\,\,\dd S
=0 ,
\end{equation}
where $\bol{n}$ is the unit outward normal to the bounding surface $\p\Omega$.

In what follows, we will see that the majority of the remaining terms in the Jacobi identity can be expressed in divergence form and will vanish when subject to the boundary conditions in equation (\ref{gen3D-functionals-boundary}).

There are 8 terms involving zero $\delta/\delta P$ and two $\delta/\delta\lr{\gmr}$, which cancel out when cyclically summed over $F$, $G$ and $H$ as follows:
\begin{align}
&\sumcyc \lrs{
\fdvdv{G}{u_i}{\bol{u}}\cdot\nabla\fdv{F}{\lr{\gmr}}
-\fdvdv{F}{u_i}{\bol{u}}\cdot\nabla\fdv{G}{\lr{\gmr}}
}\p_i\fdv{H}{\lr{\gmr}}=0
\label{terms-wo-P-start}, \\
&\sumcyc 
\fdvdv{F}{u_i}{\lr{\gmr}}\,\p_i\fdv{H}{\lr{\gmr}}\,\nabla\cdot\fdv{G}{\bol{u}}
-\lr{\nabla\cdot\fdv{H}{\bol{u}}} \fdvdv{G}{\bol{u}}{\lr{\gmr}}\cdot\nabla\fdv{F}{\lr{\gmr}}
=0
, \\
&\sumcyc 
-\fdvdv{G}{u_i}{\lr{\gmr}}\,\p_i\fdv{H}{\lr{\gmr}}\,\nabla\cdot\fdv{F}{\bol{u}}
+\lr{\nabla\cdot\fdv{H}{\bol{u}}} \fdvdv{F}{\bol{u}}{\lr{\gmr}}\cdot\nabla\fdv{G}{\lr{\gmr}}
=0
, \\
&\sumcyc \nabla\cdot\fdv{H}{\bol{u}} \lrs{
\fdvdv{F}{\lr{\gmr}}{\lr{\gmr}}\,\nabla\cdot\fdv{G}{\bol{u}}
-\fdvdv{G}{\lr{\gmr}}{\lr{\gmr}}\,\nabla\cdot\fdv{F}{\bol{u}}
} = 0
.
\end{align}

There are 9 terms involving zero $\delta/\delta P$ and one $\delta/\delta\lr{\gmr}$, which cancel out when cyclically summed over $F$, $G$ and $H$ as follows:
\begin{align}
&\sumcyc \frac{1}{\gmr}\,\epsilon_{ijk}\,\omega_i
\lr{\fdvdv{G}{u_j}{\bol{u}}\cdot\nabla\fdv{F}{\lr{\gmr}}}\fdv{H}{u_k}
-\frac{1}{\gmr}\,\bol{\omega}\cdot
\lr{\fdvdv{F}{u_i}{\bol{u}}\cp\fdv{G}{\bol{u}}}\p_i\fdv{H}{\lr{\gmr}}=0
, \\
&\sumcyc -\frac{1}{\gmr}\,\epsilon_{ijk}\,\omega_i
\lr{\fdvdv{F}{u_j}{\bol{u}}\cdot\nabla\fdv{G}{\lr{\gmr}}}\fdv{H}{u_k}
-\frac{1}{\gmr}\,\bol{\omega}\cdot
\lr{\fdv{F}{\bol{u}}\cp\fdvdv{G}{u_i}{\bol{u}}}\p_i\fdv{H}{\lr{\gmr}}=0
, \\
&\sumcyc -\frac{1}{\gmr}\,\epsilon_{ijk}\,\omega_i\,
\fdvdv{F}{u_j}{\lr{\gmr}}\lr{\nabla\cdot\fdv{G}{\bol{u}}}\fdv{H}{u_k}
-\lr{\nabla\cdot\fdv{H}{\bol{u}}}\lrs{\frac{1}{\gmr}\,\bol{\omega}\cdot\fdv{F}{\bol{u}}\cp\fdvdv{G}{\bol{u}}{\lr{\gmr}}}
=0
, \\
&\sumcyc \frac{1}{\gmr}\,\epsilon_{ijk}\,\omega_i\,
\fdvdv{G}{u_j}{\lr{\gmr}}\lr{\nabla\cdot\fdv{F}{\bol{u}}}\fdv{H}{u_k}
-\lr{\nabla\cdot\fdv{H}{\bol{u}}}\lrs{\frac{1}{\gmr}\,\bol{\omega}\cdot\fdvdv{F}{\bol{u}}{\lr{\gmr}}\cp\fdv{G}{\bol{u}}}
=0
, \\
&\lrs{\nabla\cp\lr{\frac{1}{\gmr}\fdv{F}{\bol{u}}\cp\fdv{G}{\bol{u}}}}\cdot\nabla\fdv{H}{\lr{\gmr}}
=\nabla\cdot\lrs{\fdv{H}{\lr{\gmr}}\,\nabla\cp\lr{\frac{1}{\gmr}\fdv{F}{\bol{u}}\cp\fdv{G}{\bol{u}}}}
.
\end{align}

There are 4 terms involving zero $\delta/\delta P$ and zero $\delta/\delta\lr{\gmr}$.
All terms cancel out when cyclically summed over $F$, $G$ and $H$ as follows:
\begin{align}
&\sumcyc \frac{1}{\lr{\gmr}^2}\,\epsilon_{ijk}\,\omega_i\,\bol{\omega}\cdot
\lr{\fdvdv{F}{u_j}{\bol{u}}\cp\fdv{G}{\bol{u}} + \fdv{F}{\bol{u}}\cp\fdvdv{G}{u_j}{\bol{u}}} \fdv{H}{u_k}=0
, \\
&\sumcyc \frac{1}{\gmr}\,\bol{\omega}\cdot\lrs{\nabla\cp\lr{\frac{1}{\gmr}\fdv{F}{\bol{u}}\cp\fdv{G}{\bol{u}}}}\cp\fdv{H}{\bol{u}}
+\frac{1}{\lr{\gmr}^2}\lr{\bol{\omega}\cdot\fdv{F}{\bol{u}}\cp\fdv{G}{\bol{u}}} \nabla\cdot\fdv{H}{\bol{u}}
\notag \\ & \qquad
=\sumcyc \nabla\cdot\lrc{ \bol{u}\cp\lrs{\lrc{
\lr{\frac{1}{\gmr}\fdv{G}{\bol{u}}\cdot\nabla}\fdv{F}{\bol{u}}
-\fdv{F}{\bol{u}}\cdot\nabla \lr{\frac{1}{\gmr}\fdv{G}{\bol{u}}}
+\fdv{F}{\bol{u}}\,\nabla\cdot\lr{\frac{1}{\gmr}\fdv{G}{\bol{u}}}
}\cp\frac{1}{\gmr}\fdv{H}{\bol{u}} }}
\label{Jacobi-identity-tricky-term}
.
\end{align}

The divergence form of the most complicated term (\ref{Jacobi-identity-tricky-term}) in the Jacobi identity can be obtained by referring to Lemma 4.1 in the paper \cite{Abdelhamid2015}.
Introducing the simplified notation 
\begin{gather}
\bol{F}=\fdv{F}{\bol{u}},~~~~\bol{G}=\fdv{G}{\bol{u}},~~~~\bol{H}=\fdv{H}{\bol{u}}, \\
F_i=\fdv{F}{u_i},~~~~G_i=\fdv{G}{u_i},~~~~H_i=\fdv{H}{u_i},
\end{gather}
the left-hand side of (\ref{Jacobi-identity-tricky-term}) can be expressed as
\begin{align}
&\frac{1}{\gmr}\,\bol{\omega}\cdot\lrs{\nabla\cp\lr{\frac{1}{\gmr}\fdv{F}{\bol{u}}\cp\fdv{G}{\bol{u}}}}\cp\fdv{H}{\bol{u}}
+\frac{1}{\lr{\gmr}^2}\lr{\bol{\omega}\cdot\fdv{F}{\bol{u}}\cp\fdv{G}{\bol{u}}} \nabla\cdot\fdv{H}{\bol{u}}
\notag \\ &
=\bol{\omega}\cdot\lrs{\nabla\cp\lr{\bol{F}\cp\frac{\bol{G}}{\gmr}}}\cp\frac{\bol{H}}{\gmr}
+\lr{\bol{\omega}\cdot\frac{\bol{F}}{\gmr}\times\frac{\bol{G}}{\gmr}}\nabla\cdot\bol{H}
\notag \\ &
=\bol{\omega}\cdot\lrs{
\lrc{\nabla\cp\lr{\bol{F}\cp\frac{\bol{G}}{\gmr}}}\cp\frac{\bol{H}}{\gmr}
+\lr{\frac{\bol{F}}{\gmr}\cp\frac{\bol{G}}{\gmr}}\nabla\cdot\bol{H}
}
.
\label{Jacobi-identity-divergence-term-1}
\end{align}
These terms can be transformed into
\begin{align}
&\sumcyc \bol{\omega}\cdot\lrs{
\lrc{
\lr{\frac{\bol{G}}{\gmr}\cdot\nabla}\bol{F}+\bol{F}\lr{\nabla\cdot\frac{\bol{G}}{\gmr}}
-\lr{\bol{F}\cdot\nabla}\frac{\bol{G}}{\gmr}-\frac{\bol{G}}{\gmr}\lr{\nabla\cdot\bol{F}}
}
\cp\frac{\bol{H}}{\gmr}
+\lr{\frac{\bol{F}}{\gmr}\times\frac{\bol{G}}{\gmr}}\nabla\cdot\bol{H}
}
\notag \\ &
=\sumcyc \bol{\omega}\cdot\lrs{
\lrc{
\lr{\frac{\bol{G}}{\gmr}\cdot\nabla}\bol{F}+\bol{F}\lr{\nabla\cdot\frac{\bol{G}}{\gmr}}-\lr{\bol{F}\cdot\nabla}\frac{\bol{G}}{\gmr}
}
\cp\frac{\bol{H}}{\gmr}
}
\notag \\ &
=\sumcyc \bol{u}\cdot\nabla\cp\lrs{
\lrc{
\lr{\frac{\bol{G}}{\gmr}\cdot\nabla}\bol{F}+\bol{F}\lr{\nabla\cdot\frac{\bol{G}}{\gmr}}-\lr{\bol{F}\cdot\nabla}\frac{\bol{G}}{\gmr}
}
\cp\frac{\bol{H}}{\gmr}
}
\notag \\ & \qquad\qquad\qquad\qquad
+\nabla\cdot\lrc{\bol{u}\cp\lrs{
\lrc{
\lr{\frac{\bol{G}}{\gmr}\cdot\nabla}\bol{F}+\bol{F}\lr{\nabla\cdot\frac{\bol{G}}{\gmr}}-\lr{\bol{F}\cdot\nabla}\frac{\bol{G}}{\gmr}
}
\cp\frac{\bol{H}}{\gmr}
}}
. \label{Jacobi-identity-divergence-term-2}
\end{align}
The first equality holds because the fourth and last terms cancel each other when summed over permutations.
The first term can be expressed using Levi-Civita symbol $\epsilon_{ijk}$ as
\begin{align}
&\epsilon_{ijk}u_i\p_j\lrs{\epsilon_{klm}\lr{
\frac{G_n}{\gmr}\,\p_nF_l+F_l\,\p_n\frac{G_n}{\gmr}-F_n\p_n\frac{G_l}{\gmr}
}\frac{H_m}{\gmr}}
\notag \\ &
=\lr{\delta_{il}\delta_{jm}-\delta_{im}\delta_{jl}}u_i\p_j\lrs{
\frac{G_n}{\gmr}\frac{H_m}{\gmr}\,\p_nF_l+F_l\,\frac{H_m}{\gmr}\,\p_n\frac{G_n}{\gmr}-F_n\frac{H_m}{\gmr}\,\p_n\frac{G_l}{\gmr}
}
\notag \\ &
=u_i\p_j\lrs{
\frac{G_k}{\gmr}\frac{H_j}{\gmr}\,\p_kF_i-\frac{G_k}{\gmr}\frac{H_i}{\gmr}\,\p_kF_j
+F_i\,\frac{H_j}{\gmr}\,\p_k\frac{G_k}{\gmr}-F_j\,\frac{H_i}{\gmr}\,\p_k\frac{G_k}{\gmr}
-F_k\,\frac{H_j}{\gmr}\,\p_k\frac{G_i}{\gmr}+F_k\,\frac{H_i}{\gmr}\,\p_k\frac{G_j}{\gmr}
}
\notag \\ &
=u_i\p_j\lrs{
\frac{H_j}{\gmr}\,\p_k\lr{F_i\,\frac{G_k}{\gmr}}-\frac{H_i}{\gmr}\,\p_k\lr{F_j\,\frac{G_k}{\gmr}}
-F_k\,\frac{H_j}{\gmr}\,\p_k\frac{G_i}{\gmr}+F_k\,\frac{H_i}{\gmr}\,\p_k\frac{G_j}{\gmr}
}
. \label{Jacobi-identity-intermediate}
\end{align}
The first and third terms are transformed using the Leibniz rule as
\begin{align}
\frac{H_j}{\gmr}\,\p_k\lr{F_i\,\frac{G_k}{\gmr}}
&=\p_k\lr{F_i\,\frac{G_k}{\gmr}\frac{H_j}{\gmr}}-F_i\,\frac{G_k}{\gmr}\,\p_k\frac{H_j}{\gmr}
,
\\
-F_k\,\frac{H_j}{\gmr}\,\p_k\frac{G_i}{\gmr}
&=-\p_k\lr{F_k\,\frac{G_i}{\gmr}\frac{H_j}{\gmr}}+\frac{G_i}{\gmr}\,\p_k\lr{F_k\,\frac{H_j}{\gmr}}
.
\end{align}
Equation (\ref{Jacobi-identity-intermediate}) can now be summarized as
\begin{align}
&u_i\p_j\p_k\lr{F_i\,\frac{G_k}{\gmr}\frac{H_j}{\gmr}-F_k\,\frac{G_i}{\gmr}\frac{H_j}{\gmr}}
+u_i\p_j\lrs{\frac{G_i}{\gmr}\,\p_k\lr{F_k\,\frac{H_j}{\gmr}}-\frac{H_i}{\gmr}\,\p_k\lr{F_j\,\frac{G_k}{\gmr}}}
\notag \\ & \qquad\qquad
+u_i\p_j\lr{F_k\,\frac{H_i}{\gmr}\,\p_k\frac{G_j}{\gmr}-F_i\,\frac{G_k}{\gmr}\,\p_k\frac{H_j}{\gmr}}
.
\end{align}
Each term cancels out when summed over permutations.
In particular, the first term cancels due to the commutativity of partial derivatives:
\begin{equation}
\sumcyc u_i\p_j\p_k\lr{F_i\,\frac{G_k}{\gmr}\frac{H_j}{\gmr}-F_k\,\frac{G_i}{\gmr}\frac{H_j}{\gmr}}
=\sumcyc u_i\p_j\p_k\lr{F_i\,\frac{G_j}{\gmr}\frac{H_k}{\gmr}-F_k\,\frac{G_i}{\gmr}\frac{H_j}{\gmr}}
=0
.
\end{equation}
Therefore, the left-hand side of equation (\ref{Jacobi-identity-tricky-term}) is equal to the term expressed as divergence in equation  (\ref{Jacobi-identity-divergence-term-2}), confirming that equation (\ref{Jacobi-identity-tricky-term}) holds.
The right-hand side of equation (\ref{Jacobi-identity-tricky-term}) can be transformed as follows:
\begin{align}
& \int_\Omega \nabla\cdot\lrc{
\bol{u}\cp\lrs{
\lrc{\lr{\frac{\bol{G}}{\gmr}\cdot\nabla}\bol{F}
-\bol{F}\cdot\nabla \lr{\frac{\bol{G}}{\gmr}}
+\bol{F}\,\nabla\cdot\lr{\frac{\bol{G}}{\gmr}}
}
\cp \frac{\bol{H}}{\gmr}
}} \dd^3x
\notag \\ &
= \int_{\p\Omega}
\bol{n}\cdot\bol{u}\cp\lrs{ \lrc{
\lr{\frac{\bol{G}}{\gmr}\cdot\nabla}\bol{F}
-\bol{F}\cdot\nabla \lr{\frac{\bol{G}}{\gmr}}
+\bol{F}\,\nabla\cdot\lr{\frac{\bol{G}}{\gmr}}
}
\cp \frac{\bol{H}}{\gmr}
} \dd S
\notag \\ &
= \int_{\p\Omega} \lrc{
\lr{...\cdot\bol{n}}\lr{\frac{\bol{H}}{\gmr}\cdot\bol{u}}
- \lr{...\cdot\bol{u}}\lr{\frac{\bol{H}}{\gmr}\cdot\bol{n}}
} \dd S
\notag \\ &
= \int_{\p\Omega}
\lr{\frac{\bol{H}}{\gmr}\cdot\bol{u}}
\bol{n}\cdot\lrc{
\lr{\frac{\bol{G}}{\gmr}\cdot\nabla}\bol{F}
-\bol{F}\cdot\nabla \lr{\frac{\bol{G}}{\gmr}}
+\bol{F}\,\nabla\cdot\lr{\frac{\bol{G}}{\gmr}}
} \dd S
\notag \\ &
= \int_{\p\Omega}
\lr{\frac{\bol{H}}{\gmr}\cdot\bol{u}}
\bol{n}\cdot\nabla\cp\lr{\frac{\bol{F}\cp\bol{G}}{\gmr}} \dd S .
\end{align}
Since $\bol{F}\cdot\bol{n}=\bol{G}\cdot\bol{n}=0$, 
\begin{equation}
\frac{\bol{F}\cp\bol{G}}{\gmr}=\lambda\,\bol{n}
\end{equation}
with a scalar function $\lambda$.
Since $\Omega$ is smoothly bounded, there locally exist scalar functions $C$ and $\lambda'$ such that $\bol{n}=\lambda'\nabla C$, leading to the conclusion that
\begin{equation}
\bol{n}\cdot\nabla\cp\lr{\frac{\bol{F}\cp\bol{G}}{\gmr}}
=\lambda'\nabla C\cdot\nabla\lr{\lambda\lambda'}\cp\nabla C=0.
\end{equation}
Thus, the right-hand side of equation (\ref{Jacobi-identity-tricky-term}) vanishes.

\subsection{Proof that the Poisson bracket (\ref{Poisson-bracket-3D}) satisfies the Jacobi identity}
\label{sub: Proof that Poisson Bracket (Poisson-bracket-3D) Satisfies Jacobi Identity}

In this subsection, we demonstrate that the Poisson bracket (\ref{Poisson-bracket-3D}) satisfies the Jacobi identity.
Observe again that the divergence term appearing in the following expression 
\begin{equation}
\lr{\nabla\cp\delta\bol{u}}\cdot\frac{1}{\gmr}
\frac{\delta F}{\delta\bol{u}}\times\frac{\delta G}{\delta\bol{u}}
=\nabla\cdot\lrs{\frac{1}{\gmr}\,\delta\bol{u}\cp\lr{\frac{\delta F}{\delta\bol{u}}\times\frac{\delta G}{\delta\bol{u}}}}
+\delta\bol{u}\cdot\nabla\cp\lr{\frac{1}{\gmr}\frac{\delta F}{\delta\bol{u}}\times\frac{\delta G}{\delta\bol{u}}}
\label{curl-delta-u2},
\end{equation}
identically vanishes under the boundary condition \eqref{3D-functionals-boundary}. Indeed, 
the divergence term in equation \eqref{curl-delta-u2} vanishes because $\fdv{F}{\bol{u}}\cp\fdv{G}{\bol{u}}$ is parallel to $\bol{n}$, which is a consequence of the fact that both $\fdv{F}{\bol{u}}$ and $\fdv{G}{\bol{u}}$ are perpendicular to $\bol{n}$.

Thus, the functional derivatives of $\lrc{F,G}$ are 
\begin{align}
\frac{\delta}{\delta u_i}\lrc{F,G}
&=\lrs{\nabla\cp\lr{\frac{1}{\gmr}\fdv{F}{\bol{u}}\cp\fdv{G}{\bol{u}}}}_i
+\frac{1}{\gmr}\,\bol{\omega}\cdot
\lr{\fdvdv{F}{u_i}{\bol{u}}\cp\fdv{G}{\bol{u}} + \fdv{F}{\bol{u}}\cp\fdvdv{G}{u_i}{\bol{u}}}
\notag\\ &\qquad
+\fdvdv{G}{u_i}{\bol{u}}\cdot\nabla\fdv{F}{\lr{\gmr}}
-\fdvdv{F}{u_i}{\lr{\gmr}}\,\nabla\cdot\fdv{G}{\bol{u}}
-\fdvdv{F}{u_i}{\bol{u}}\cdot\nabla\fdv{G}{\lr{\gmr}}
+\fdvdv{G}{u_i}{\lr{\gmr}}\,\nabla\cdot\fdv{F}{\bol{u}}
,
\end{align}
\begin{align}
\frac{\delta}{\delta\lr{\gmr}}\lrc{F,G}
&=-\frac{1}{\lr{\gmr}^2}\,\bol{\omega}\cdot\fdv{F}{\bol{u}}\cp\fdv{G}{\bol{u}}
+\frac{1}{\gmr}\,\bol{\omega}\cdot\fdvdv{F}{\bol{u}}{\lr{\gmr}}\cp\fdv{G}{\bol{u}}
+\frac{1}{\gmr}\,\bol{\omega}\cdot\fdv{F}{\bol{u}}\cp\fdvdv{G}{\bol{u}}{\lr{\gmr}}
\notag\\ &\qquad
+\fdvdv{G}{\bol{u}}{\lr{\gmr}}\cdot\nabla\fdv{F}{\lr{\gmr}}
-\fdvdv{F}{\lr{\gmr}}{\lr{\gmr}}\,\nabla\cdot\fdv{G}{\bol{u}}
\notag\\ &\qquad
-\fdvdv{F}{\bol{u}}{\lr{\gmr}}\cdot\nabla\fdv{G}{\lr{\gmr}}
+\fdvdv{G}{\lr{\gmr}}{\lr{\gmr}}\,\nabla\cdot\fdv{F}{\bol{u}}
.
\end{align}
There are $7\cdot2+7=21$ terms in $\lrc{\lrc{F,G},H}$.
These terms are exactly the same terms appearing in equations \eqref{terms-wo-P-start}--\eqref{Jacobi-identity-tricky-term}, and cancel out under the same logic.

\end{document}